# Classification of Supersecondary Structures in Proteins Using the Automated Protein Structure Analysis Method

Sushilee Ranganathan [1], Dmitry Izotov [1], Elfi Kraka [1], and Dieter Cremer *[1,2]

[1]*Department of Chemistry, University of the Pacific, 3601 Pacific Avenue, Stockton, CA 95211, USA,* [2] *Department of Physics, University of the Pacific, 3601 Pacific Avenue, Stockton, CA 95211, USA.*

*(E-mail: dcremer@pacific.edu)*

**Abstract**: The Automated Protein Structure Analysis (APSA) method is used for the classification of supersecondary structures. Basis for the classification is the encoding of three-dimensional (3D) residue conformations into a 16-letter code (3D-1D projection). It is shown that the letter code of the protein makes it possible to reconstruct its overall shape without ambiguity (1D-3D translation). Accordingly, the letter code is used for the development of classification rules that distinguish supersecondary structures by the properties of their turns and the orientation of the flanking helix or strand structures. The orientations of turn and flanking structures are collected in an octant system that helps to specify 196 supersecondary groups for αα–, αβ–, βα–, ββ-class. 391 protein chains leading to 2499 supersecondary structures were analyzed. Frequently occurring supersecondary structures are identified with the help of the octant classification system and explained on the basis of their letter and classification codes.

## 1. Introduction

The richness of protein structure is parallel to the richness of protein functionality. However, the encoding of protein functionality in its structure is so complex and dynamic that there exists no generally applicable "procedure" to derive functionality from structure. Inherent to this problem is the fact that the description, analysis, and understanding of protein structure and the assessment of the various (un)folding modes are, despite numerous achievements in this field, still at its infancy. [1-10] The analysis and understanding of protein structure still remains the first hurdle on the way of assessing protein functionality. [2] Hence, there is need for a rapid and systematic, computer-based analysis of a large number of proteins of different structures that provides a platform for the categorization and understanding of structure and functionality.

One entry point to the understanding of protein structure is the analysis of the backbone conformation. Once the backbone structure is understood residue by residue, side chains can be added to the backbone and conformationally assessed. The residue dihedral angles $\phi$ and $\psi$ give a local, discrete, and bond-wise assessment of the backbone. Since early investigations [4], it has been described how these values "vary considerably in irregular secondary structures", especially β-strands, and how they "can be subject to considerable experimental error" as they heavily depend on the orientation of peptide groups. The same investigation [4] also remarks on how an "averaging-out of error" is needed by considering a group of adjacent residues. Nevertheless, numerous studies, that are general [5, 6] as well as detailed [2, 7], have still been carried out on structure assessment of proteins up to the supersecondary level using a $\phi$–$\psi$-based analysis method. Detailed articles [8] that review the attempts made by various groups give valuable information on the progress in this field. The formation of supersecondary and tertiary structure from suitable fragments collected in structure libraries [9, 10] and a piece-wise construction of proteins from tertiary units has been suggested.

Recently, we have presented a new Automated Protein Structure Analysis (APSA) method that can encode the conformational features of the protein backbone into a 16-letter code that combines to *words* representing secondary structural units, *phrases* (combination of words connected by short letter strings that correspond to turns) representing supersecondary structural units, and *sentences* corresponding to protein chains. [11-13] Each residue in a protein is conformationally described by the 16-letter code, however this description differs in so far from the normal description of residues in terms of dihedral angles $\phi$ and $\psi$ as it is based on a coarse-grained representation of the backbone. It focuses on non-local conformational features reflecting the interplay between vicinal residues when analyzing the three-dimensional (3D) shape of the protein backbone. The 16 letters are the tools for the primary encoding of the 3D structure of the backbone into a one-dimensional (1D) string of characters. The primary code contains all conformational details and is the starting point for the secondary encoding, which groups the residues into helices, strands, and turns. [13] These secondary units are used for describing tertiary protein structure.

In this work, we use the primary and secondary encoding of APSA for the description and classification of supersecondary structures. For this purpose, we will introduce the 3D-1D-3D projection-translation method (3D-1D-3D PTM), which is an extension of the previously introduced 3D-1D projection method. [13] The latter has the task to condense 3D conformations into a 1D code, whereas the former uses the 1D code and translates it into 3D supersecondary structure information. Such a 1D-3D translation will be an indispensable prerequisite if the APSA encoding is used for any classification of supersecondary and tertiary protein structure. If for example the 1D-3D translation leads to ambiguities in the overall shape of a protein or any of its substructures, then the letter code (primary or secondary) cannot be used for a classification system of protein structure that is based on conformational features alone. Hence, before an APSA classification method is launched one has to first verify the validity and applicability of 3D-1D-3D PTM, which will be one objective of the current work (Section 3).

3D-1D-3D PTM will guide us in developing simple rules based on the encoding of the backbone in terms of the 16 conformational letters to describe, recognize, and classify supersecondary structure (Section 4). The classification of supersecondary structures will be detailed and will lead to a large number of groups (Section 4) where group identification labels collect information about the conformational composition of turn and flanking helices (strands) and how the latter are oriented relative to each other. Considering these orientations, an octant system will be invented to merge supersecondary groups to larger groups associated with space regions of the octants (Section 5). The classification system will be applied to 249 proteins and 2499 supersecondary structures to demonstrate its general applicability, high degree of automation, and usefulness in understanding protein structure.

**2. Computational Methods**

The mathematical basis of APSA for obtaining a smooth backbone line of a protein and calculating curvature and torsion diagrams as a function of the arc length *s* of the backbone line, κ(s) and τ(s), has been amply discussed in our previous publications. [11,12] Therefore, we just mention here that the $C_\alpha$ atoms of the protein are suitable anchor points for calculating a smooth backbone line with the help of a spline. By this approach, a coarse-grained backbone structure is obtained, which does no longer provide individual directions of the backbone bonds (and by this the backbone angles ϕ and ψ), but accurately contains the shape of the protein backbone and all its structural units from the secondary to the tertiary level. Each protein backbone can be represented by its κ(s) and τ(s) diagrams that reveal typical patterns for specific conformational features. Ideal highly regular and less regular conformations of secondary structure can be identified by setting up ranges (*windows*) of κ(s) and τ(s) values. [11] This makes it possible to specify body, termini, entry, and exit residues for natural helices and ß-strands. [12] Also, a systematic description of distortions, kinks, and breaks in secondary structure becomes possible. [12]

Additional work revealed that that the torsion τ is sufficient to distinguish between secondary structural units. [13] The torsion parameter identifies the chirality of any helical or extended structure (positive τ values: right-handed twist; negative τ values: left-handed twist) and sensitively reflects all conformational changes along the protein backbone. The one-dimensional torsion diagrams can be easily replaced by a (continuous) sequence of τ–*windows* where each window is associated with a residue (given by its $C_\alpha$ position) and represented by a letter. [13] This is possible because each residue of the protein backbone possesses a typical conformation associated with a typical τ-peak, τ-trough or, in rare cases, a τ-base (a flat segment of the backbone line with τ-values close to zero) that can be translated into a simple 16-letter code.

For the purpose of rapidly testing the results of the classification system introduced in this work, we developed in this work a computer routine that determines the orientation of a helix by calculating its axis. The helical axis can be obtained by adjusting a cylinder to the helix. This problem implies the minimization of the following penalty function $F$

$$F = \frac{1}{2}\sum_{i}(N_i^2 - R^2) \tag{1}$$

$$N_i^2 = M_i^2 - (\overrightarrow{M_i}, \vec{a})^2 \tag{2}$$

$$\overrightarrow{M_i} = \vec{P_i} - \vec{O} \tag{3}$$

$$\vec{a} = \{\cos\alpha\sin\beta, \sin\alpha\sin\beta, \cos\beta\} \tag{4}$$

where $\vec{a}$ is the unit helix axis vector with $\alpha$ and $\beta$ being polar angles, $\vec{O}$ is an arbitrary point on the helix axis, $\vec{P}_i$ denotes helix points, and $R$ is the cylinder radius. Minimization has to be done with respect to six variables, i.e. the three coordinates of point $\vec{O}$, the orientation angles $\alpha$ and $\beta$ of the helix axis, and the radius $R$ of the cylinder. This problem can be solved numerically using as a standard optimization method the conjugate gradient method [14]. The first derivatives of the functional required by the latter method are then calculated analytically.

The functional of form (1) gives the sums of the squared distances of the helix points from the cylinder surface. In case of an ideal circular helix, the minimum of function F is zero. In reality, however, it is a positive number close to zero because of the non-ideal geometry of protein helices. The minimum point obtained is also sensitive to the choice of the initial guess and the number of helical points. Even in the case of an ideal helix, functional (1) has multiple minima if the number of helical points is less than five (assuming four amino acids per turn in case of an α-helix). In this case, there can be a functional minimum erroneously indicating a direction perpendicular to the helix axis. We fine-tuned the program to avoid these situations by using five helix points and the vector connecting the first and the fifth helix atom as a suitable initial guess for the helix axis. The geometric center of a tetrahedron spanned by the first four atoms of the helix was used as the initial guess for the axis point $\vec{O}$. Five points at both sides of each helix were used to detect the axis directions.

Two sets of proteins will be used in this investigation. The first one comprises 143 protein chains taken from the PDB [15] and is a combination of a set of 77 proteins [12] and 93 proteins [13] (excluding dublicates) used in the earlier works. The second set is a collection of 249 chains obtained from a list of proteins generated by PDB-REPRDB. [16] For all proteins considered, the structural resolution is < 2.0 Å. Proteins with a sequence agreement of 20% or more with any other member of the set were excluded to guarantee a large spread in protein structures. The individual members of the 143- and 249-set are listed in the Supporting Information.

Encoding of the protein structure, classification procedures for secondary and supersecondary structures, and the calculation of the helix orientations in αα-structures are carried out by the program APSA. [17]

## 3. The 3D-1D-3D Projection-Translation Method

Table 1 lists some key terms used for the 3D-1D projection method that is the starting point for the development of 3D-1D-3D PTM. Each residue and its conformation are presented by a torsion peak (trough) in the τ(s) diagram. Peaks and troughs reflect two characteristic torsion properties (shape and sign), which are summarized by two-character symbols. The first character (referring to the shape of the torsion peak) indicates whether a given residue possesses a stretched (e for extended; possible characters: B, J) or coiled conformation (L for looping; possible characters: A, H, N, U, W, 3). The second character indicates the twisting of the residue conformation (+: right-handed; -: left-handed, Table 1) as reflected by the sign of τ(s) peak (trough). In this connection it has to be mentioned that because of the coarse-grained description of the backbone a single τ-peak (trough) always includes information on the conformation of neighboring residues (the *conformational environment*) and therefore identifies a piece of a helical coil, which possesses the property of handedness. Each letter represents the conformational region of a residue and a string of letters connects all conformational regions (and associated residues) to the overall shape of the backbone segment. Starting from a given letter string, the overall shape of the backbone can always be constructed by translation of the two-character symbols into orientations of the backbone line in 3D. In a previous investigation, we have already shown that the overall shape of supersecondary structures (expressed by the form of the connecting turn and the relative orientations of turn and flanking structures 1 and 2) can be accurately represented in the letter code. [13] Accordingly, it can also be retrieved from such a code, which is the topic of this article. The 1D-3D translation step is illustrated with the help of some representative examples shown in Figure 1.

Table 1, Figure 1

We partition a supersecondary structure into structure 1 (helix or strand), turn, and structure 2 (helix or strand). Aligning the axis of structure 1 along the z-axis in the way that the turn starts next to the origin of the coordinate system, the orientation of the turn can be assessed from the letter code of the exit residue X (Table 1) of structure 1 and those of the turn residues. Similarly, the letter codes of the turn, the entry residue N for structure 2 (which is part of the turn) and starter residue S (Table 1) of structure 2 define its orientation.

In Figure 1, superimposed supersecondary structures of the αα- (Figure 1a-1c, 1g), αβ- (Figure 1d), βα- (Figure 1e), and ββ-type (Figure 1f) are shown in form of ribbon diagrams. For each turn of the αα-structures, the letter code is given from the exit residue of structure 1 through the starter residue of structure 2 with the turn codes given in parentheses. For structures containing strands, neither the exit not the starter residue conformation is required. Strand recognition, by APSA requires that its exit and starter residue to be a member of the extended (B, J) conformations. Unless specified, the stands are always left-handed, so the exit/starter is either B- or J- and therefore not mentioned. For the purpose of simplifying the comparison of the different supersecondary structures shown in one diagram, all structure 1 ribbons are superimposed in such a way that their exit residue $C_\alpha$ atom exactly coincide and the orientation of structure 1 is about the same. The turn and structure 2 take their respective orientations, the conformations of which are specified by the letter codes.

*Exit influences on the bending of a turn*: In Figure 1a, two αα-supersecondary structures from 3TS1 and 1HMD are connected by turns of the same (B- B-) conformation starting at L130 and Q66, respectively. The *helix 1 exit residues* (given by X in Figure 1a) are both extended but of opposite sign (B+ and B- respectively), which causes the turns to be oriented in different directions. If one considers the $C_\alpha$ atoms of exit (helix 1) and starter residue (helix 2) in 3TS1 and 1HMD to form a plane (indicated by a dashed triangle in Figure 1a), the orientation of this plane with respect to the axis of *helix 1* is determined by the conformations of the exit residues as reflected by the shape property of the τ-peak and the corresponding letter code. A transition from the A+ conformation of the helix into the B conformation of the exit (and vice versa at the starter of *helix 2*) leads to approximately perpendicular arrangement between plane and the axis of helix 1 (helix 2). The turns of 3TS1 and 1HMD bend differently in this plane at X yielding turn orientations that differ by 80-90°. The bending orientation of the turns away from the direction of helix 1 is for 3TS1 the result of a transition from a right-handed helix (A+) into a right-handed (B+) extended conformation according to .. A+; X = B+ (B-; B-) S=_3; A+ .. and for 1HMD a farther bending results due to the transition from A+ to left handed B- according to .. A+; X = B- (B-; B-) S=_A; A+ . Since the turn residues agree in shape (B) and sign (-), and the starter of helix 2 has either _A (1HMD) or _3 (3TS1) character (similar transitions from a left-handed turn into a right-handed helix), helix 2 of 1HMD and 3TS1 have comparable orientations to the superimposed helix 1 structures. They are not exactly parallel due to difference in starters, but point in an upward direction relative to the XSS plane drawn in Figure 1a. We call these orientations to be qualitatively perpendicular to the plane XSS (Figure 1a) and by this parallel to each other thus specifying two orientation angles for each structure 2 (one with reference to the plane; one with reference to the axis of structure 1).

*The five orientation parameters define the overall shape of a supersecondary structure.:* There is no need to specify the exact orientations of turn and structure 2. Instead, the orientations need only to be specified in a relative sense semiquantitatively when two or more supersecondary structures are compared.  For this purpose, the axes of helices (strands) 1 from different supersecondary structures are coincided. It is meaningful to define planes that contain (largely) the turns to be compared. This can be done (as in the case of Figure 1a) by just specifying the (superimposed structure 1) exits X and the two (or more) starter $C_\alpha$ points S of structures 2. With the help of such a plane and the torsions of the X and S residues, we can define five geometrical parameters that determine the orientations of turn and structure 2.

i) *Tilting of the plane*: The plane specified can be orthogonal or tilted with regard to the axis of helix 1, influenced by the e/L conformation of the exit; e. g., an ‘A+’ to ‘e’ transition leads approximately to a perpendicular arrangement between plane and axis 1 (Figure 1a).  - ii) In-plane *bending of the turn*: An opposite sign at the exit residues of two supersecondary structures compared (e.g. B+ vs B-) is predominantly responsible for the bending of the turns within the plane. Tilting of the reference plane and bending of the turns define the relative turn orientations. - iii) *Turn Length*: For the determination of the orientations of structures 2, one has to shift an imaginary coordinate system from the exit of structure 1 to the starter residue of starter 2, i.e. in the direction of the turn by a distance equal to its length. The number of residues gives the length of the turn, which is of primary consideration for the pattern of codes, however represents only an auxiliary parameter when determining the orientation of structure 2. - iv) Tilting of axis 2: The sign of the entry residue is predominantly responsible for the out-of-plane tilting of helix 2. If the sign does not change from turn to structure 2, then the axis of structure 2 will not deviate much from the plane; if it does, axis 2 is more or less perpendicular to the plane. – *v) Relative orientation of axis 1 and axis 2*: General terms are used for describing relative orientation of 2 helix axes as being “parallel”, “perpendicular”, etc., which can be more quantitatively described by the dihedral angle (axis 1; turn axis; axis 2). In the primary code, a change from L+ (in helix 1) to <e> (extended conformation of exit and turn residues) and then back to L+ conformations in helix 2 (L+-<e>-L+ pattern) is predominantly responsible for a parallel orientation. However, this holds only if the sign combination in <e> leads to an approximately zero net in-plane rotation of the turn, e.g. when <e> = B-; B-, and the starter S = _A. In a similar way, rules can be given for other relative orientations of axis 1 and axis 2 such as an orthogonal one.

*Entry turn influences and the orientation of helix 2*: In Figure 1b, segments of an αα-supersecondary structure from 1GOX (green; 1-residue turn at G319: .. A+; X=B- (A-) S=W+; A+ ..) and 1CDP (brown; 1-residue turn at N7: .. A+; X=B- (B-) S=_A; A+ ..) are shown that have the same exit B-. The turn residues (equal to the *helix 2 entries*) differ by their τ-shape property being A- (looping peak) for 1GOX and B- (extended peak) for 1CDP. Correspondingly, the two turns have the same *length* (1 residue) and not much difference in *bending* when projected onto the XSS plane (each turn has a negative torsion corresponding to a left-handed coiling; Figure 1b). However, the difference between an ‘A-’ and a ‘B-’ turn residue is that the $C_\alpha$ point of the former is forced into looping conformation below the plane whereas that of the latter is close to it. Corresponding to this, helix 2 of 1GOX is out-of the plane oriented in almost a V-shape to helix 1, whereas for 1CDP, helix 2 forms with helix 1 an ‘L’. A change from a looping (A) to an extended (B) conformation at the entry of structure 2 determines the difference in the orientations of their axis as shown for 1GOX and 1CDP. This difference can be quantified within the limits of the τ-ranges specified for the letter code.

*Increase of turn bending in αα-structures*: The maximum difference in bending is obtained when the opposite bending given by a change from a plus to a minus-torsion (in the plane defined in Figure 1a) is augmented by switching from a looping to an extended conformation, e.g. from a helical A+ to an extended B- conformation. This is shown for 1HMD (purple; .. A+ X=B- (B-; B-) S=_A; A+..; turn starting at Q66) superimposed on its own *helix 1* (blue; …A+A+A+…) as shown in Figure 1c. The helix pattern gives the most coiled structure whereas the turn adopts the most extended (or uncoiled) form, i.e. α-helix and β-strand are compared. When another structure from 1COL with exit and turn having B+ conformation (green; .. A+; X=B+ (B+; B+; J-; B-) S=_A; A+ ..; turn starting at L165) is superposed in the same way, its turn orientation (up to the J-) falls in between the orientations of the two other structures (Figure 1c). This is because this first half of the turn is extended, but with positive torsion τ. The orientation of helix 2 for 1COL is parallel to that of helix 1 because reversion of the turn direction by a J- and following B- residue both being of the 'e' type [13] brings helix 2 in a parallel direction (Figure 1c).

*Simultaneous tilting and bending of αβ-structures:* The inferences drawn from αα-supersecondary structures are applicable to the αβ-, βα-, and ββ-structures as well (Figures 1d-1g). Figure 1d shows *helix 1* superimposed for four αβ-structures taken from 1CDP (brown; turn start at A72; pattern: ... A+; X = B- (_A; B+) β2...), 1GYM (purple; R155; pattern: ... A+; X = B+ (N+; B+) β2...), 2FOX (blue; E46; pattern: ... A+; X = B+ (B-; W+) β2...), and 1GOX (green; G120; pattern: ... A+; X = B+ (W-; B+) β2...). The latter 3 structures have positive exits (X = B+) and can be seen to coil in the same right-handed fashion as *helix 1* up to that point. The same 3 structures also have similar bending within the indicated reference plane (the turn of 1GOX has to be projected onto the plane). The 1CDP structure is different, and never enters the plane, as can be predicted form the B- exit. The turns of 1GOX, 2FOX and 1GYM lie below the plane (L- = W-), on the plane (e- = B-), and bent the other way (L+ = N+), respectively. The β strands of 1CDP and 1GOX are equally parallel to the helix axis, but positioned to its left and right. This is guided by the B- and B+ exit followed by L+ = _A) and L- = W- conformations and by a common B+ pattern. The W- and _A rotation, though opposite, is not equal, introducing a slight disparity of ≈15°. In 1GYM, the B+ follows an N+ and gives a planar obtuse angle that leads to an orientation close to the reference plane.

*Helix distortions in βα-structures:* The conformation of the starter (rather than the entry residue, which is still in the turn) is responsible for "inserting" the turn into helix 2. The presence of distortions immediately following the starter residue in helix 2 will lead to a different orientation of the rest of structure *2*, though the first few residues match. Figure 1f shows four superimposed βα-supersecondary structures from protein 1ABE having the same exit and a 0-residue turn, i.e. strand 1 directly enters helix 2. In the three examples 1ABE 66 (green; P66; ... β1; S=_A; B-; A+…), 1ABE 231 (brown; I231; ... β1; S=_3; B-; A+…), and 1ABE 204 (yellow; M204; ... β1; S=_A; B-; A+…), there is the common pattern β; S; B-; h; h. Hence, there is a distortion after the start of helix 2 at B which becomes obvious when compared with the undistorted pattern β; S; h; h; h. These three examples coincide well with respect to the orientations of their helix 2 structures. The fourth structure (1ABE 110; purple; A110; ... β1; S=W+; A+; A+… corresponding to ß; S; h; h) posses a regular *helix 2*. For this one reason, the orientation of helix 2 in 1ABE 110 does not match those of the other three helix 2 structures, the difference in orientation being ≈90° as caused by a switch from B- to A+ (Figure 1e).

*Simultaneous tilting and bending in ββ-structures*: The ββ-examples (Figure 1f) with 3-residue turns taken from 1CYO (purple; L25; ... β1; (N-; W-; B+) β2…) and 1GYM (brown; H82; ... β1; (W-; N-; B+) β2…) have as a common pattern β1; (L-; L-; B+) β2 where 'β' indicates the flanking β-strands and L- negative looping members. Figure 1f reveals that these are oriented toward the same direction. The presence of two left-handed looping members in the turn shows progress in a left-helical fashion, up to the point of sign change at the third turn residue. In 1GYM (blue; M121; ... β1; (B-; _A; B+) β2…), the sign change appears earlier (at the second turn residue) leading to right-handed bending. The following strand is approximately at 90° to the first because one B- and one B+ cancel each other by rotation and the net effect of the turn is an _A entry into the conformation of strand β2. This, we know, is a 90° bending in the turn plane (Figure 1f). The orientation of ß2 in 1ABE (green: G247; ... β1; (B+; B+; W-) β2…) is totally different from those of the three other ß2-structures; the first B+ imparts a 90° rotation in the reference plane of *strand 1* and the second B+ implies an out-of-plane bending of 90°. *Strand 2* is oriented in between a parallel and a perpendicular structure because of the W- looping conformation.

*Trends in sign of turn influence bending:* Figure 1g depicts four superimposed helix 1 structures (schematically represented by a single helix) that disembogue in four 2-resifue turns (residues T1 and T2) that are differently oriented (orientations O1, O2, O3, O4) due to a different sign of the helix exit X. Four different sign patterns are considered for the string X; T1; T2, namely O1 (...A+; X+; T1+; T2+); O2 (...A+; X-; T1-; T2-); O3 (...A+; X-; T1+; T2+); O4 (...A+; X+; T1-; T2-). O1 and O4 share the same exit sign (X+) and accordingly, bend forward, along the right handed direction (Figure 1g). However, O1 points further down in approximately the same direction (T1+; T2+) whereas O4 bends upward from T1 onwards (T1-; T2-). Orientations O2 and O3 share an opposite exit pattern (X-) leading to a left-handed twist toward the back. They follow O1 and O4 in the way that the (T1+; T2+) combination leads again to an upward, the (T1-; T2-) combination to a downward direction. In this way, Figure 1g reveals that an inversion of the X sign or the whole sign pattern for X; T1; T2 leads to change of the orientations O as indicated in Figure 1g by black (change of X sign) and purple arrows (inversion of sign pattern).

1D-3D translations can be given for each part of the letter code with the result that the overall shape of the protein backbone emerges from its 1D representation in terms of strings of letters. The translation step is unique since different letter codes lead to different non-contradictory shapes of supersecondary structures. Once this aspect has been clarified, there is actually no longer any need to carry out the 1D-3D translation. If one would want to know the relative orientations of structures 1 and 2 as well as that of the turn, then these could be much easier (and more accurately) calculated in form of vectors directly taken from the Cartesian coordinates of a protein. However, the verification of the 1D-3D translation is a necessary prerequisite for the classification of supersecondary structures based on primary and secondary encoding as the main objective of this work (see section 4.

## 4. Classification of Supersecondary Structures

In the previous section it has been demonstrated that the 3D shape of a supersecondary structure can be assessed from the string code. Accordingly, it must be also possible to carry out a geometrically based classification of supersecondary structures according to the overall shape by utilizing just the primary and secondary encoding of the protein backbone in terms of conformational letters based on the torsion diagrams. The general strategy for such a classification is detailed in Figure 2. This strategy follows the description of turns by considering their lengths, bending, and tilting (Section 3). Armed with these aspects of the 1D-3D translation, one can organize the flow of information as shown in Figure 2. This contains the shape and sign property information of the turn segment and the secondary structure residues neighboring the turn. The information is organized into a hierarchy that groups structures with increasing detail at each level. The data is collected for each supersecondary structure in a protein from both the primary and secondary code.

Figure 2

The first step of the classification proposed involves an initial sorting of regions according to their letter codes and the identification of *flanking secondary structures* that are connected by turns to αα-, αβ-, βα-, and ββ- supersecondarty structures. It is noteworthy that the ends of the turn take the conformation of the adjacent secondary structures. Therefore, an understanding of the turn structures implies the analysis of the turn flanks. Next, the *number of turn residues* is specified (Figure 2). As discussed in connection with a bending of a turn, turns with the same overall orientation are grouped by their sign change. *Homogeneous turns* refer to those without any sign changes and can be *all-positive* or *all- negative*. Alternatively, the turns can have a *mixture* of signs. Both these groups are then classified based on whether the exit is positive or negative (Figure 2).

A group containing structures with turns having sign changes is not comparable to the *all-negative turn* or *all-positive turn* groups because the former includes more variable structures even after further classification by the sign of the exit. An additional criterion for sign-mixtures is based on whether the turn contains all the positive peaks in one half and all the 'negative peaks' (= troughs) in the other where such a pattern identifies the group of *segregated turns*. This group splits up in the +/- segregated subgroup (minus signs follow positive signs) and the -/+ segregated subgroup (plus signs following minus signs; Figure 2). Alternatively, the signs can alternate (Figure 2).

Considering all possibilities for a turn with sign mixture, a total of six different sign groups, all at the same level of classification, emerge for this case, which are comparable with the four *homogeneous turn groups* as indicated by the same coloring of the boxes in Figure 2. At the next level of classification, the presence of looping peaks is denoted and at the final level the sign of the entry (the last residue of the turn) to structure 2 of the supersecondary unit is used as classification criterion.

**Setting up of group identifications (group-ID)**: Because of the large number of groups, an automated labeling procedure is invented that represents the conformation of the region being described in terms of all the information collected in the flowchart of Figure 2. The labeling strategy represented by the character string F.#.M.X.O.L.N.R.U.S is described in Table 2. This is the supersecondary level grouping and is determined by APSA as an additional classification, after the primary and the secondary codes have been given. The categorization into the major classes of αα, αβ, βα and ββ is denoted at the F level (F for flanking structures) by the starting character 1, 2, 3, 4. It follows the number (#) of residues in the turn, the sign changes in the turn (M), the sign of the exit (X) residue of structure 1, the order (O) of signs in the turn, the position of the first looping (L) residue in the turn, the sign of the entry (N) residue into structure 2, the regularity (R) of structures 1 and/or 2, unwinding (U) of structures 1 and/or 2, and finally the sign (S) of the ß-strand if structure 2 is a ß-strand (Table 2).

Table 2

It is superfluous to note down the sign of the starter residue of a helix because it is always positive. Accordingly, the starter information is omitted for αα- and βα-structures. Other examples where different parts of the labeling become interdependent are given in Table 2. For example, a mixture of sign changes can occur only if the turn contains 2 or more residues (comments for 'M' in Table 2). The order of sign change will be applicable only if there *is* a sign change, i.e. if the turn is a mixture (comments for 'O' in Table 2). The sign of entry into the β-strand is not as dominant in influencing orientation as the τ-sign of the strand itself and so this information is discounted under 'N' for αβ- and ββ-structure (Table 2). Instead, for these same groups of structures, the sign of the starter is included (this reflects the sign of the β-strand; comments for 'S', Table 2).

The regularity index ('R', Table 2) is determined by the presence/absence of distortions (symbol 'AD' in the secondary code, Table 1) in the helices (considering the first 6 helical residues) flanking the turn. Strands are stretched; the translation of 4 residues in a strand conformation is greater than in a helix conformation. So the regularity of strands is not restricted to the presence of distortions in a particular number of residues as done in helices. In the secondary code, a 'B' stands for regular strand whereas a 'b' denotes a kinked one.

For the classification purpose, not all details of the primary and secondary letter code (Table 1) have to be included. As the comments for 'L' and 'U' indicate (Table 2), the group-ID contains indices for the mere presence of looping peaks or the negative helix unwinding peaks, respectively. The exact position of occurrence and conformation can always be determined from the primary code.

Figure 3

**Classification hierarchy**. Figure 3 gives the classification hierarchy and schematically lists the number of groups possible at a given levels. Starting from 4 classes (αα, αβ, βα, ββ) and considering a 4-residue turn, a maximum of 480 groups is obtained. The analysis of the peak codes of the exit (structure 1), turn and entry (structure 2) results in the large number of groups that are detailed in the analysis and indexing. This is the level reached at the end of Figure 2. We then analyze (by the example in Figure 1f) the regularity of the ends of helices and the β-strands to obtain even more specified groups. This corresponds to the labeling in Table 2. The results at this level are expected to provide maximum detail with regard to the backbone conformation of a protein. Starting from the level of maximum detail, one can start merging similar groups to larger meaningful groups by dropping the dependence on criteria that increase detail. Hence, the strategy presented by Figure 3 provides a sophisticated basis to derive useful classification systems for supersecondary structures with less conformational detail.

The classification system F.#.M.X.O.L.N.R.U.S of Table 2 was implemented into the computer program APSA and applied to a dataset of 143 chains of 143 proteins (see Supporting Information). Results are summarized in Tables 3 and 4, which list the most strongly populated groups and present some statistics on the distribution of 2499 supersecondary structural examples investigated. A list of the top 100 most strongly populated groups sorted by frequency and the conformational group IDs according to the F.#.M.X.O.L.N.R.U.S key is given in Table 3. These 100 groups represent supersecondary structures that occur at least 5 times. The four most frequent groups among all belong to the βα-class (F = 3) with no turn residue (# = 0) and negative exits (X = -) indicating left-handed ß-strands. Among them, 102 examples have regular structural flanks (R=0), 83 have irregular strands (F = 2, 3, 4; R = 1, 2), 78 have irregular helices (F = 1, 2, 3; R = 1, 2), and 65 have both helices and strands irregular (R = 3, Table 2). This result corresponds to the observation that the most common entry into a helix is the β-strand conformation [18].

The next most frequently occurring supersecondary group is from the ββ-class with one (# = 1) positive (M = +) extended (L = 0) turn residue where both the β-strands are kinked (R = 3) (52 examples with 4.1.+.-.*.0.*.3.0.-; Table 3). Note that kinks of structure are not classified to match in 3D. As mentioned earlier, the exact conformation of the irregularity can be obtained from the primary code. The same conformation as above with just the first strand kinked (4.1.+.-.*.0.*.1.0.-) occurs 48 times (Table 3). The third set of most frequent ββ-structures (occurring 42 times) has two right-handed turn residues where the first turn member is looping and the second β-strand is kinked.

The largest group among αβ-structure (49 members, Table 3) is one with no turn and a negative extended helix exit. Both flanks are found to be distorted among this group. All β-strands occurring among the groups discussed so far are left-handed. Only a small percentage of strands is right-handed. The most frequently occurring αα-structure has 15 examples with positive helix exit followed by a 2-residue, extended, all negative turn (pattern: α (e+; e-; e-) α).

The total of 2499 supersecondary structural examples investigated led to 745 groups (Table 4) that are distributed similarly among the four classes αα, αβ, βα and ββ, however, the αα-class covered the least (375) number of examples. This tallies with the largest αα-group having only 15 examples and 117 out of 179 groups contain only 1 example (Table 4). This wide distribution of conformations further highlights the need for a merging step indicated in Figure 3. Though the longest turns of each class is found to have over 10 residues in it, the much shorter average length (about 2 residues) shows the strong disfavor for long turns.

The detailed labeling (Table 2) was done to convert individual letters of the character code into well-formed phrases [13] that provide a detailed conformational representation of supersecondary structure (without explicit 3D representations and comparisons) and that makes further manipulation of protein structure easy. The merging step indicated in Figure 3 is also to be based on the character code. The aim of this step is to show that a particular region in conformational space is populated by a specific supersecondary structure identified by its APSA code. For the purpose of illustrating some objectives of the merging step we compare in Figure 4 turns of three αα-structures with similar bending wherein one (grey color) with different bending but the same length stands out. We can simplify the comparison by saying that all four turns are located approximately in a common (average) plane where the direction of the fourth turn deviates by about 45° from the directions of the first three turns. All four turn directions can be placed in a quadrant defined by the first three turn directions and a perpendicular axis and in this way the four turns form a new unified group defined by the quadrant.

A fifth turn of different length (purple, Figure 4) would also fall into the new group provided the criterion of turn length would be dropped. Clearly, the maximum spread of structures with small turns is not as great as the maximum spread that can be obtained from structures whose turns are longer. Hence, by deleting criteria like turn direction and turn length that increase the number groups, a drastically reduced number of groups can be obtained. The examples of Figure 4 reveal that such a simplification has also its drawbacks because 4 of the 5 turn directions have to be considered as border cases enclosing the maximally possible bending angle of 90°. A useful classification system based on a merging of more detailed groups has to consider this, which will be discussed in the following section.

## 5. Condensing the Number of Groups by Using Octants and Cones

In section 3, supersecondary structures were compared by aligning the axis of structure 1 with the z-axis. Also, we have shown that the character code specifies in which direction the turn and the axis of structure 2 point. We can collect these directions in the eight octants of the coordinate system where however this has to be done in two steps, first with regard to the turn directions and then with regard to the axis directions of structure 2. In Figure 5a, a x,y,z-coordinate system is sketched and its eight octants are schematically indicated by numbers 1 to 8. Octant 1 is enclosed by the +x-, +y-, and +z-axis. Octants 1 through 4 are labeled in a clockwise fashion as viewed from the +y-axis and octants 5 through 8 lie 'behind' them (Figure 5a).

As for structure 1, its exit (i.e. the $C_\alpha$ atom of the exit residue) is positioned at the origin of the coordinate system; the axis of structure 1 (in the case of Figure 5a, a helix) is parallel to the +z-axis and slightly shifted in –x direction to accommodate the exit X at the origin (Figure 5a). The direction of viewing of the coordinate system is along the +x-axis facing the exit of structure 1. Thus arranged, one can specify to which of the octants the turn vector points. For determining the direction of the axis of structure 2, the coordinate system has to be moved along the turn vector to its end point keeping all axes parallel to the first coordinate system. The definition of all octants is kept and one can allocate the direction of axis 2 to an octant in the same way as done for the turn vector.

Each octant corresponds to manifold of turn (structure 2) directions, which are associated with many different supersecondary structures. The group-IDs are constructed at a level of great detail. They will not meaningfully converge into common patterns when collected into octants without further specifying additional regions within each octant that they can occupy. Thus, a system of circular cones is introduced to distinguish structures that lie close to the x-, y-, or z-axes (Figure 6b). The structures that do not fall into cones are described by the octant numbers highlighted by a star as in '8*'. Each of the two orientation vectors that have to be described are denoted by the octant number and the axis of the cone they belong to. For example, "5x6y" refers to a turn that lies along the +x-axis of octant 5 (only +x is possible for octant 5) and the second helix vector lies along the -y-axis of octant 6 (only –y is possible for octant 6).

We use circular cones with an aperture of 2 x 30° (non-touching cones) and 2 x 45° (touching cones), which fill the octant spaces increasingly and leaving smaller inter-cone spaces. Each circular cone (Figure 6b) is "quartered" by the two planes that define the octants. The maximum number of theoretically possible orientations is given by the total number of regions contained in each octant (3 cone + 1 inter-cone = 4) times the number of octants (8) squared (for the octant system of turn and structure 2), which is 1,024 in total. This is not a small number and nothing is gained by this new system, given that the primary code based classification leads to 748 classes (Figure 3). If however, each quarter of a cone along the same axis is combined, then there will be 6 cones (along the 6 axes) and the inter-cone spaces will be described by the 8 octants, thus giving 14 possible orientations for each of the two octant systems and 196 (14 x 14) in total.

The cone partitioning of the octant spaces is shown in Figure 6c. Since structure 1 is aligned along the z-axis, it is reasonable to identify the four perpendicular directions by easy to remember geographical terms rather than the coordinate axes. The cone along the +x-axis (splitting up into 1x, 4x, 5x, 8x) points to the south (S), the cone along the –x-axis to the north (N), the cone along the +y-axis (1y, 2y, 3y, 4y) to the east (E), and the cone along the –y-axis (5y, 6y, 7y, 8y) to the west (W) direction. The +z- (1z, 2z, 5z, 6z) and –z-cones (3z, 4z, 7z, 8z) are denoted as up (U) and down (D). The "5x6y" example translates with this notation to "SW" and one can immediately visualize that the turn points south and the helix west (with orientation of helix 1 along the U-direction). The inter-cone labels do not change (as in '8*').

A dataset of 391 chains (corresponding to 391 proteins; see Supporting Information) was analyzed using the octant-system for the classification of supersecondary structures. In Figure 5d, the results of the analysis of 504 αα-supersecondary structures are summarized by showing the populations of the various space regions of the octant system. The orientations of structure 2 relative to that of structure 1 (aligned along the U-direction) are given. The 30° and 45° cone populations are differentiated for the purpose of providing additional insight into the occupation of space regions close to the bisectors of the coordinate system. The assignment of axis directions in supersecondary structures as derived from the primary and secondary encoding of the protein backbone was verified in a large number of cases by an auxiliary program that determines helix and turn directions from available protein coordinates [15] and in a smaller number of cases by direct inspection of computer models of a protein rotated in 3D-space. Hence, the data presented in Figure 5d (see also Table 5) has been verified.

The largest number of examples (133) is found in the U direction, i.e. the axes of helix 1 and helix 2 are close to being parallel as is typical for hairpin structures. This correlates well with the large percentage of extended turn conformations (e = B, J: 73 %) shown in Table 3, entry 12. They keep the turn straight, whereas the transitions A+ to e and e to A+ guarantee the 90° orientation changes from helix 1 to turn and then to helix 2 (see Figure 1a and relevant text).

These observations explain why hairpins are the most frequent αα-structures found in proteins. [19]

When using Efimov's description of supersecondary structures, an 'L-shape' [18] would include all forms with the axis of helix 2 pointing to the N, S, E, and W cones thus leading to 203 supersecondary structures. From the starter of *helix 2,* the observed directional preference for the cones over the interspaces suggests a preference of 90° orientations manifested by the conformation of the starter of helix 2 (transition from e to L; see above). However, one cannot attribute this directional preference to a single factor alone as the axis direction of helix 2 is determined by the combined influence of the turn conformation as well. - E, W, and N direction are equally populated (41 – 44 structures) whereas the S-direction with 74 examples is much stronger occupied. This is a direct consequence of the positioning of helix 1 (Figure 5a) with its exit X coinciding with the origin of the coordinate system and the helix 1 axis being parallel to the +z-axis shifted in the direction of –x (N-direction). If the turn develops straight out of helix without any energy costing sideward distortion it must point to S and the same direction is preferred for the turn-helix 2 transition. Accordingly, the S-direction is strongly populated.

Figure 5d reveals that an increase in the cone size from 30 to 45° strongly reduces the populations of the inter-cone spaces. The most drastic change is visible in 1* (from 48 to 4). The change is complemented by an increase in the populations of the S-cone (from 42 to 74), U-cone, (from 92 to 133), and E-cone (from 23 to 44). Clearly, there is a continuum of possible conformations and helix 2 orientations, and any classification that divides space into regions is arbitrary in its boundaries. On the other hand, any classification that makes it possible to recognize individual supersecondary structures as belonging to one or the other group is useful and provides a basis for further classification. In the case of αα-supersecondary structures we proceed by using the 45° cones because the edges of the cones touch each other and the cone inter-space is secluded from the inter-spaces of neighboring octants.

Table 5 lists 22 groups extracted from a set of 504 αα-structures investigated by considering just 1-residue turns. For each group, the octant regions into which the turn and the axis of helix 2 point are both given. Examples of supersecondary structures with large overall similarity have been merged so that a larger space region can be given, such as '8' (helix 2 axis orientation, group 1). Some commonly used literature names ('L'; 'V' [18]) are also given in Table 5 to express in a symbolic way how the two helices of an αα-supersecondary structure are oriented with respect to each other. Clearly, these terms are vague; an 'L' can have an infinite number of orientations for helix 2 in the NSEW plane of the octant system where an L-structure pointing E is conformationally different from one that is pointing W.

There is a strong correlation between group characteristics (as expressed by the group ID) and space regions of the octant system, which confirms that the primary encoding and subsequent group labeling translates into structure orientations in 3D. As one moves down the column containing the group-IDs (Table 5), for a change in the 'R' (regularity of the flanking structures) and 'U' (depending on R) parameters, the same space orientation is kept. For example, given that the space region 'W' is partly contained in octant '5', the presence or absence of irregularity in helix 1 has no influence on the orientation of the turn vector. This independence is because the vector is constructed for any type of helix (α or not) placed along the +z-axis, with its exit at the origin. So the following turn vector and the helix 2- axis vector are judged from that point onwards, independent of earlier conformations.

## 6. Conclusions

Any simplification of protein structure must lead to a loss of details where these details may not be relevant for a description of the global shape of the protein. In previous work, we have developed APSA [11-13] to the point that the 3D structure of a protein can be projected into a 1D string of letters. This involves the use of torsion ranges and the loss of excessive detail. Therefore it was an essential question of this investigation whether the overall shape of a protein and its most important substructures can be reconstructed from its letter code. By developing the 1D-3D translation method, we have demonstrated that secondary, supersecondary, and total protein structure can be reconstructed from its letter code without any ambiguity. Essential elements of the translation are classifying residue conformations into extended and looping forms in a right and left-handed sense, and identifying those residues that influence orientations and overall shape of supersecondary structure. The transition from *structure 1* into the turn is influenced by the exit residue and from the turn into *structure 2* by the entry and starter residue conformations. A few rules discussed in this and previous work establish the 3D-1D-3D projection-translation method that guarantees a meaningful representation of the overall shape of protein structure starting from the 16-letter code.

The 3D-1D-3D PTM is the prerequisite for the development of the F.#.M.X.O.L.N.R.U.S classification key (Table 2) for supersecondary structures. Analysis of regions from 143 proteins yields 749 groups with different supersecondary structures. These groups are merged by bringing together closely oriented structures in space with the help of an octant system to map the orientations of turn and *structure 2* relative to that of structure 1. The octant regions can be further partitioned by introducing six circular cones of defined aperture along the coordinate axes. In this way, a total of 14 space regions that lead to 196 groups were defined for each of the four supersecondary classes ($\alpha\alpha$, $\alpha\beta$, $\beta\alpha$, $\beta\beta$). The analysis of 391 proteins led to 504 $\alpha\alpha$ structures that were collected in groups, according to their overall shape, using the octant system. The $\alpha\alpha$-hairpin was identified as the most frequently occurring structure (133 examples; 26 %). The frequency of a number of typical supersecondary structures was similarly explained utilizing the letter code and the 3D-1D-3D PTM protocol.

We have demonstrated here for the first time how APSA provides the starting point for a rapid and meaningful classification of tertiary protein structure. In future work we will apply APSA to structure-based studies such as predication and folding of proteins.

**Supporting Information**: PDB identification numbers of all proteins investigated are given in the Supporting Information. Also, curvature and torsion diagrams, $\kappa(s)$ and $\tau(s)$, and the backbone line are shown for each protein.

**Acknowledgements**

DC and EK thank the University of the Pacific for support. Support by the NSF under grant CHE 071893 is acknowledged

## Figure Captions

**Figure 1.** Ribbon representations of supersecondary structures (structure 1, turn, structure 2) superimposed for structure 1 (exact superimposition of exit X of helix 1 (strand 1) to determine the orientation of turn and/or that of structure 2 relative to the orientation of structure 1 via the letter code. Starter residue S of structure 2 and axis (strand or helix) are given in some cases. **a**) αα-Structures: 1HMD (yellow, turn starting at 66) and 3TS1 (purple, turn starting at 130). Dashed lines indicate the XSS plane. **b**) αα-Structures: 1GOX (green, turn starting at 319) and 1CDP (brown, turn starting at 7). The XSS plane is indicated by dashed lines. **c**) αα-Structures: 1HMD (purple, turn starts at 66) and 1COL (green, turn starts at 165). The direction of helix 1 is given by superimposing the helix of 1HMD (blue) onto itself. The XJS plane is given by dashed lines. **d**) αß-Structures: 1GOX (green, start of turn at 120), 1CDP (brown, start of turn at 72), 1GYM (purple, start of turn at 155), and 2FOX (blue, start of turn at 46). The XWB plane is indicated by dashed lines. **e**) βα-Structures: 1ABE 66 (green), 1ABE 110 (purple), 1ABE 204 (yellow) and 1ABE 231 (brown). Numbers after PDB-ID indicate turn start. **f**) ßß-Structures: 1GYM (blue, turn starting at 82), 1CYO (purple, turn starting at 25), 1GYM (brown, turn

starting at 121), and 1ABE (green, turn starting at 247) are superimposed for the last 3 residues of strand ß1. The 3 residues define a plane (indicated by dashed lines) containing the ß1-kink of 1GYM (brown). **g**) Schematic drawing of 4 superimposed helix 1 structures with four different sign patterns thus causing the 2-residue turn (residues T1 and T2) to adopt four different directions: O1 (down, front direction) for X+; T1+; T2+; O2 (down, back direction) for X-; T1-; T2-; O3 (top, back direction) for X-; T1+; T2+; O4 (top, front direction) for X+; T1-; T2-.

**Figure 2**. Schematic flowchart (from left to right) for applying the rules of Table 2. Decision boxes in red indicate the same level of 3D information.

**Figure 3.** Schematic representation of steps 1 to 5 for the classification of supersecondary structures. Step 1 implies the classification into the αα–, αβ–, βα–, ββ-classes. Step 2 classifies according to the residue conformations by using the F.#.M.X.O.L.N rules of Table 2. Step 3 considers in addition the regularity of flanking structures as described in Table 2. Step 4 merges groups by using the octant system. Step 5 applies topology and other criteria to further reduce the number of supersecondary groups. – The insert shows how the number of groups increases (theoretically) in step 2 with the number of residues in the turn.

**Figure 4**. Ribbon representations of supersecondary structures (helix 1, turn, helix 2) superimposed for helix 1 (viewing direction along axis of helix 1). Three αα-structures (2CTS 131:yellow; 1COL 165: green; 1CDP 21: brown) possessing the same turn length and similar turn directions (helix 2 directions) are compared with a fourth αα-structure (1HMD 66: grey) having the same turn length, but a turn direction ca. 45° rotated with regard to the former turn directions. If a fifth αα-structure (1COL 49: purple) is considered with a longer turn and a 90° orientation relative to the first 3 turns, differences between the first four structures seem to become smaller on the new scale introduced by the fifth structure.

**Figure 5**. **a**) Schematic representation of the division of 3D space by the octant system (encircled numbers 1 to 8) spanned by the coordinate system. An αα-supersecondary structure (helix 1, turn, helix 2) is oriented by aligning the axis of helix 1 with the +z-axis (exit X at origin; helix 1 axis also shifted along–x). The direction of turn and helix 2 are given by red vectors. – **b**) Partitioning of the octant spaces by six circular cones oriented along the coordinate axes in plus and minus directions. The planes that are the boundaries for the octants divide the cones in four

equal parts. – **c**) Assigning geographical terms (S = south, N = north, E = east, W = west, U = up, D = down) to the cone (coordinate axis) directions. – **d**) The inter-cone space is identified by the octant number given in a) and a star (e.g., 8*). For clarification reasons, an octahedron is used where the triangle areas are connected with the inter-cone spaces according to the octant numbering of a) and the direction labeling of c). Octant triangles in the front are identified by labels outside the octahedron, which are connected with an arrow to the triangle in question. Octant triangles in the back contain the label inside the triangle. – **c**) and **d**): The number of αα-structures (obtained via the classification of 391 proteins) falling into a given octant region are shown next to each space region label where the upper numbers are obtained for circular cones with an aperture of 2 x 30° and the lower number for cones with an aperture of 2 x 45°.

**Table 1**. APSA terms needed for the classification of supersecondary structures.

| APSA Terminology | Definition/Explanation |
|---|---|
| Backbone line of a protein | Line obtained by fitting $C_\alpha$ positions with a cubic spline |
| Coarse-grained description of protein backbone | Each residue is represented by just one anchor point chosen to be $C_\alpha$ |
| κ(s)-τ(s) diagrams | Backbone line is given by curvature κ and torsion τ expressed as a function of arc length *s* |
| τ-Peaks, τ -troughs | Conformation of each residue is characterized by torsion peaks (troughs) with specific shape and sign property |
| Properties of τ-peaks (troughs) | Each segment of the backbone line from one $C_\alpha$ to the next has shape property and sign property (of torsion): 1) Shape property: Looping (L) or extended (e); 2) Sign property: right handed (+), left-handed (-)<br><br>L: A, 3, H, N, U, W; e: B, J [13] |
| Primary code | 16 letters encode all peaks (troughs) of the τ(s) diagram translating 3D shape of the backbone line into 1D character string |
| Secondary code | Second level of encoding that assigns secondary structure to predefined patterns in the primary code (h, b, AD [13]) |
| Starter of a secondary structure | The first residue starting a helix (strand): this is half into and half outside the helix (strand) |
| Exit of a secondary structure | The last residue in the helix (strand) that leads into the adjoining loop or turn segment; responsible for orienting its direction |
| Entry of a secondary structure | The residue preceding the starter which orients the secondary structure |

**Table 2**. Rules for labeling supersecondary structures.

| Criteria | Symbol | Comments | Indices used and their significance |
|---|---|---|---|
| Flanking secondary structures | F | | 1. αα<br>2. αβ<br>3. βα<br>4. ββ |
| Number of turn residues | # | | {0, 1, 2, …} |
| Whether turn contains sign changes | M | Not applicable if the turn has only 0 or 1 residue (i.e., # < 2) | *: Not applicable<br>-: homogeneously negative<br>+: homogeneously positive<br>M: mixed |
| The sign of the exit | X | | -: negative<br>+: positive |
| Order of sign changes in the turn | O | Not applicable if M is homogeneous | 0. Not applicable<br>1. Alternating<br>2. Segregated with + peaks followed by - troughs<br>3. Segregated with + peaks followed by - troughs |
| Position of the first looping residue in the turn | L | Primary code contains further details | X: exit is looping<br>0: no looping members<br>{1,2,3… }: position in the turn |
| Sign of entry (into the second structure) | N | Only for the second structure being helix; Not applicable if β-strand follows | *: Not applicable<br>-: negative<br>+: positive |
| Regularity of secondary structure (presence of distorting residues) | R | Checked for 6 helical residues bordering turns and for the entire β-strand. | 0. Regular<br>1. Structure 1 is irregular<br>2. Structure 2 is irregular<br>3. Both structures are irregular |
| Helix unwinding (presence of negative torsion in the irregular part of helix) | U | Additional details of irregularity in the absence of unwinding are present in the primary code. | 0. No helix or no unwinding in the helix<br>1. Helix 1 is unwound<br>2. Helix 2 is unwound<br>3. Both helices are unwound |
| Sign of the following β-strand (for αβ- and ββ- structures) as given by the starter | S | Sign of the preceding β-strand is given by the exit. | *: Not applicable for αα, ßα<br>-: negative sign<br>+: positive sign |

**Table 3**. The top 100 supersecondary groups that occur 5 or more times when classifying 143 proteins. [a]

| f | group-ID (F.#.M.X.O.L.N.R.U.S) | f | group-ID (F.#.M.X.O.L.N.R.U.S) | f | group-ID (F.#.M.X.O.L.N.R.U.S) |
|---|---|---|---|---|---|
| 102 | 3.0.*.-.*.0.*.0.0.* | 11 | 2.0.*.-.*.0.*.2.0.- | 6 | 2.1.+.+.*.0.*.3.0.- |
| 83 | 3.0.*.-.*.0.*.1.0.* | 11 | 4.0.*.-.*.0.*.3.0.+ | 6 | 3.1.+.-.*.0.*.2.0.* |
| 78 | 3.0.*.-.*.0.*.2.0.* | 11 | 4.1.-.-.*.1.*.1.0.- | 6 | 4.0.*.-.*.0.*.0.0.+ |
| 65 | 3.0.*.-.*.0.*.3.0.* | 11 | 4.2.M.-.3.1.*.0.0.- | 6 | 4.0.*.+.*.0.*.1.0.- |
| 52 | 4.1.+.-.*.0.*.3.0.- | 11 | 4.3.M.-.1.1.*.2.0.- | 6 | 4.1.-.-.*.1.*.3.0.- |
| 49 | 2.0.*.-.*.0.*.3.0.- | 10 | 1.1.-.+.*.0.*.1.0.* | 6 | 4.1.+.-.*.1.*.1.0.- |
| 48 | 4.1.+.-.*.0.*.1.0.- | 10 | 4.2.+.-.0.0.*.2.0.- | 6 | 4.2.+.-.0.0.*.1.0.- |
| 42 | 4.2.+.-.0.1.*.2.0.- | 10 | 4.2.M.-.3.1.*.2.0.- | 6 | 4.2.+.-.0.0.*.3.0.- |
| 38 | 2.0.*.+.*.0.*.3.0.- | 10 | 4.3.M.-.1.1.*.0.0.- | 6 | 4.6.M.-.1.1.*.2.0.- |
| 37 | 2.1.+.-.*.0.*.1.0.- | 9 | 1.2.+.-.0.0.+.0.0.* | 5 | 1.1.-.-.*.0.*.3.0.* |
| 35 | 2.0.*.-.*.0.*.0.0.- | 9 | 1.2.+.-.0.0.+.1.0.* | 5 | 1.1.+.+.*.1.*.2.0.* |
| 30 | 2.0.*.-.*.0.*.1.0.- | 9 | 3.0.*.+.*.0.*.3.2.* | 5 | 1.2.M.+.3.2.+.0.0.* |
| 30 | 2.0.*.+.*.0.*.0.0.- | 9 | 4.0.*.-.*.0.*.1.0.+ | 5 | 1.3.M.-.2.0.-.1.0.* |
| 30 | 3.0.*.-.*.0.*.2.2.* | 8 | 1.1.-.+.*.0.*.2.0.* | 5 | 1.6.M.+.1.3.-.0.0.* |
| 29 | 4.1.+.-.*.1.*.3.0.- | 8 | 2.1.+.-.*.0.*.2.0.- | 5 | 2.0.*.-.*.0.*.3.0.+ |
| 26 | 4.0.*.+.*.0.*.3.0.- | 8 | 3.0.*.+.*.0.*.1.0.* | 5 | 2.0.*.+.*.0.*.3.0.+ |
| 26 | 4.2.+.-.0.1.*.3.0.- | 8 | 3.1.+.-.*.0.*.3.2.* | 5 | 2.2.M.-.3.1.*.2.0.- |
| 23 | 3.0.*.-.*.0.*.3.2.* | 8 | 3.2.M.-.2.0.-.1.0.* | 5 | 3.0.*.+.*.0.*.0.0.* |
| 22 | 2.0.*.+.*.0.*.1.0.- | 8 | 4.1.+.+.*.1.*.3.0.+ | 5 | 3.1.-.+.*.0.*.0.0.* |
| 22 | 4.2.+.-.0.1.*.0.0.- | 8 | 4.5.M.-.1.1.*.3.0.- | 5 | 3.1.+.-.*.0.*.2.2.* |
| 22 | 4.2.+.-.0.1.*.1.0.- | 8 | 4.6.M.-.1.1.*.3.0.- | 5 | 3.1.+.-.*.0.*.3.0.* |
| 20 | 2.1.+.-.*.0.*.0.0.- | 7 | 1.1.+.-.*.0.*.3.0.* | 5 | 3.2.+.-.0.1.+.1.0.* |
| 18 | 2.0.*.-.*.0.*.3.1.- | 7 | 1.2.-.+.0.0.-.2.0.* | 5 | 3.3.M.-.2.0.-.3.0.* |
| 18 | 2.0.*.+.*.0.*.3.1.- | 7 | 1.3.M.-.2.0.-.0.0.* | 5 | 3.5.M.-.1.1.-.3.0.* |
| 18 | 2.1.+.-.*.0.*.3.0.- | 7 | 1.3.M.-.2.0.-.2.0.* | 5 | 4.1.+.-.*.1.*.2.0.+ |
| 16 | 4.0.*.-.*.0.*.2.0.+ | 7 | 2.2.+.-.0.2.*.1.0.- | 5 | 4.1.+.-.*.1.*.3.0.+ |
| 16 | 4.0.*.+.*.0.*.0.0.- | 7 | 2.2.M.-.3.1.*.3.0.- | 5 | 4.2.+.-.0.2.*.0.0.- |
| 16 | 4.0.*.+.*.0.*.2.0.- | 7 | 4.2.M.-.3.1.*.3.0.- | 5 | 4.2.+.-.0.2.*.1.0.- |
| 15 | 1.2.-.+.0.0.-.0.0.* | 7 | 4.3.+.-.0.1.*.1.0.- | 5 | 4.3.+.-.0.1.*.3.0.- |
| 15 | 4.2.+.-.0.2.*.2.0.- | 7 | 4.3.M.-.1.1.*.1.0.- | 5 | 4.3.M.-.1.1.*.3.0.- |
| 14 | 1.2.-.+.0.0.-.1.0.* | 6 | 1.1.-.-.*.0.*.0.0.* | 5 | 4.3.M.-.3.1.*.0.0.- |
| 14 | 2.0.*.-.*.0.*.1.0.+ | 6 | 1.1.-.+.*.0.*.3.0.* | 5 | 4.3.M.-.3.1.*.1.0.- |
| 13 | 3.4.M.-.1.1.-.1.0.* | 6 | 1.1.+.-.*.0.*.1.0.* | | |
| 11 | 1.1.-.-.*.0.*.1.0.* | 6 | 2.0.*.-.*.0.*.0.0.+ | | |

[a] Frequency (f) and group identification (ID) according to the F.#.M.X.O.L.N.R.U.S rules of Table 2.

**Table 4.** Statistical analysis using APSA's supersecondary structure classification procedure for a set of 143 protein chains.

| # | Parameter | Class | | | |
|---|---|---|---|---|---|
| | | αα | αβ | βα | ββ |
| 1. | Total number of groups | 179 | 198 | 161 | 207 |
| 2. | No. of entries in top 100-group sorted by frequency [a] | 19 | 20 | 20 | 41 |
| 3. | Total number of examples | 375 | 648 | 682 | 794 |
| 4. | No. of examples in top 100-group sorted by frequency [a] | 147 | 389 | 468 | 540 |
| 5. | No. of examples in the largest group | 15 | 49 | 102 | 52 |
| 6. | No. of groups with just one example | 117 | 118 | 91 | 108 |
| 7. | Average length of turn | 2.65 | 2.56 | 1.40 | 2.51 |
| 8. | Maximum length of turn | 16 | 19 | 12 | 17 |
| 9. | Frequency of the longest turn | 1 | 1 | 1 | 1 |
| 10. | Percentage of helices with negative exit among all helices | 55.73 | 66.67 | NA | NA |
| 11. | Percentage of helices with negative entry among all helices | 45.06 | NA | 20.23 | NA |
| 12. | Percentage of looping (extended) members in the turn | 27 (73) | | | |

[a] Only groups containing at least 5 examples are considered. – NA means not applicable.

**Table 5**. List of selected supersecondary groups of the αα-class, their classification code, their orientation in the octant system, and their relationship to other classification systems. [a]

| # | Group-IDs (F.#.M.X.O.L.N.R.U.S) | Orientations | | Total number | Common names [b] |
|---|---|---|---|---|---|
| | | Turn orientation | Helix 2 axis orientation | | |
| 1. | 1.1.-.-.*.0.*.0.0.* | W | 8 | 5 | (L) |
| 2. | 1.1.-.-.*.0.*.1.0.* | 5 | 1 | 19 | V |
| 3. | 1.1.-.-.*.0.*.2.0.* | | | | |
| 4. | 1.1.-.-.*.0.*.3.0.* | | | | |
| 5. | 1.1.-.-.*.X.*.1.1.* | N | 6* | 2 | L / V |
| 6. | 1.1.-.-.*.X.*.2.0.* | | | | |
| 7. | 1.1.-.+.*.0.*.0.0.* | 7 | 3 | 3 | (L) |
| 8. | 1.1.-.+.*.0.*.1.0.* | 8 | 8 | 19 | (L) |
| 9. | 1.1.-.+.*.0.*.1.1.* | | | | |
| 10. | 1.1.-.+.*.0.*.2.0.* | | | | |
| 11. | 1.1.-.+.*.0.*.2.2.* | | | | |
| 12. | 1.1.-.+.*.0.*.3.0.* | | | | |
| 13. | 1.1.-.+.*.0.*.3.2.* | | | | |
| 14. | 1.1.-.+.*.1.*.2.0.* | D | S | 3 | L |
| 15. | 1.1.+.-.*.0.*.0.0.* | W | 1 | 10 | V |
| 16. | 1.1.+.-.*.0.*.1.0.* | | | | |
| 17. | 1.1.+.-.*.0.*.1.0.* | | | | |
| 18. | 1.1.+.-.*.0.*.1.1.* | | | | |
| 19. | 1.1.+.-.*.0.*.2.2.* | | | | |
| 20. | 1.1.+.-.*.0.*.3.0.* | | | | |
| 21. | 1.1.+.-.*.1.*.1.0.* | N | N | 3 | L |
| 22. | 1.1.+.-.*.X.*.1.0.* | | | | |

[a] From a dataset of 249 chains leading to 504 αα-supersecondary structures. Group-IDs as in Table 2. See Figure 5 for definition of orientations within the octant system. Circular cones with 2 x 45° aperture are exclusively considered. Similar structures are merged to obtain simplified space regions. Parentheses denote that a subset of structures complies with the literature definition. - [b] Ref. [18] – Terms from literature discount the continuum of possible structures; in some cases more than one name has to be specified.

Figure 1a, b, c, d, e

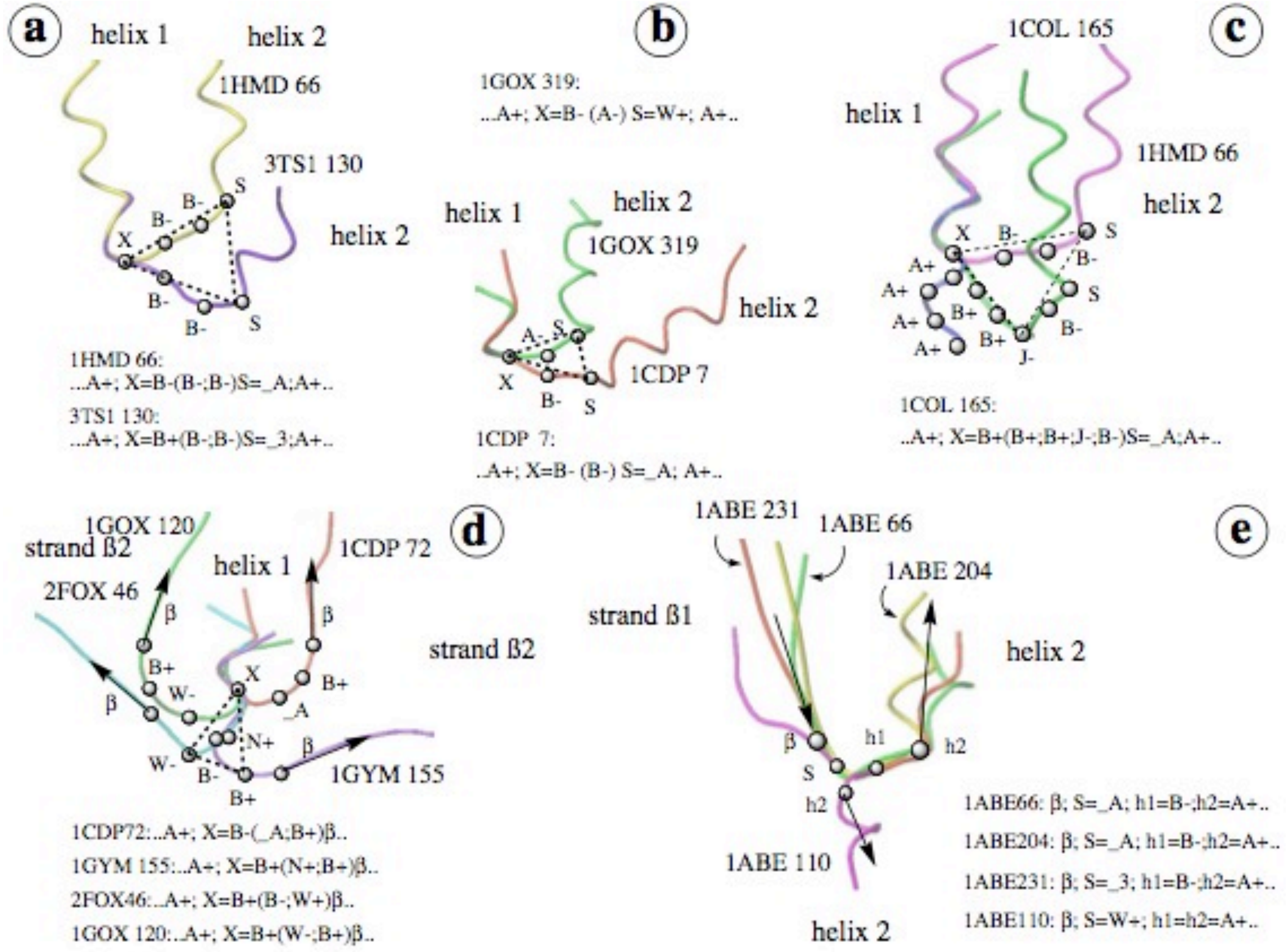

Figure 1f, g

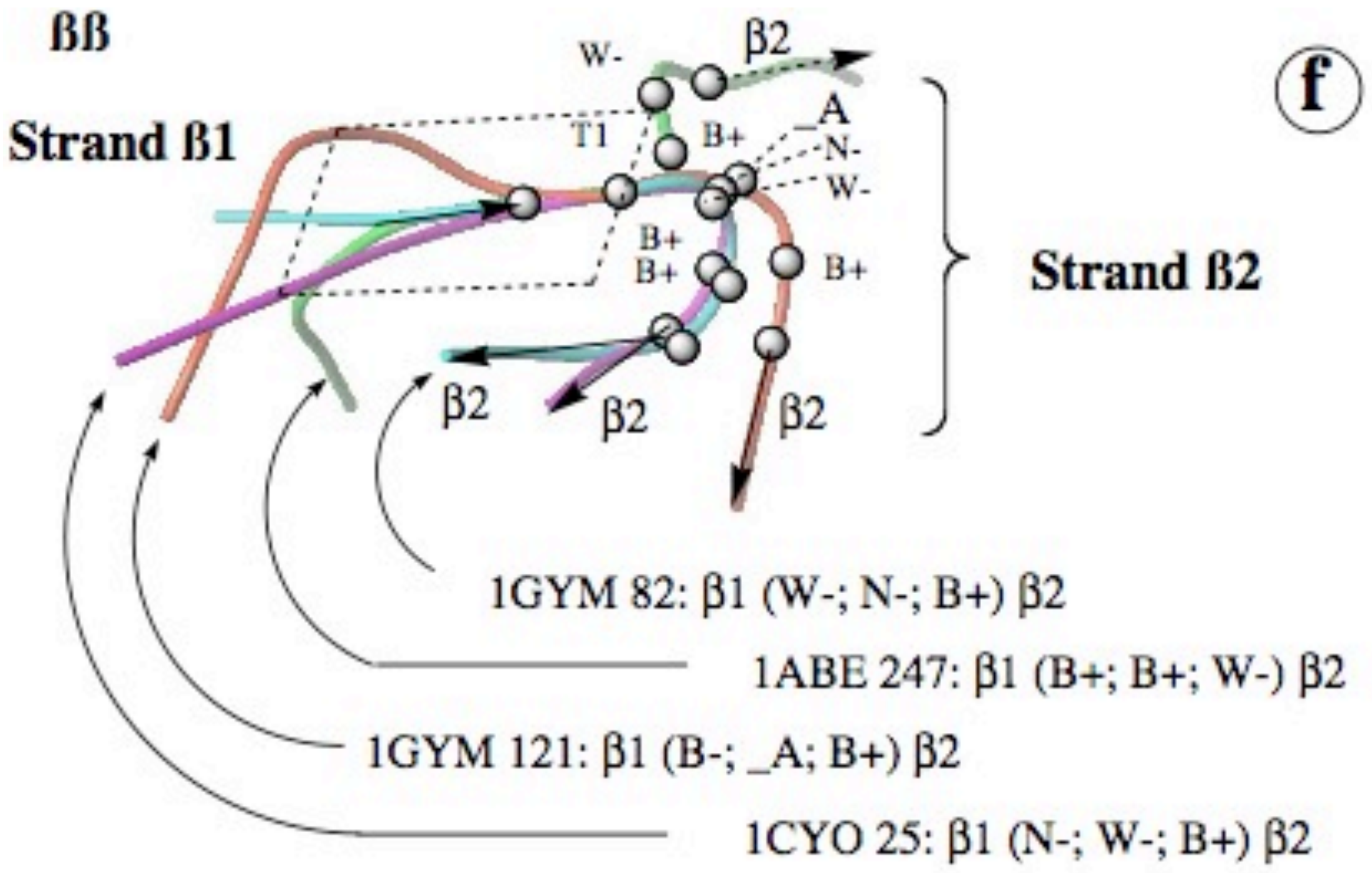

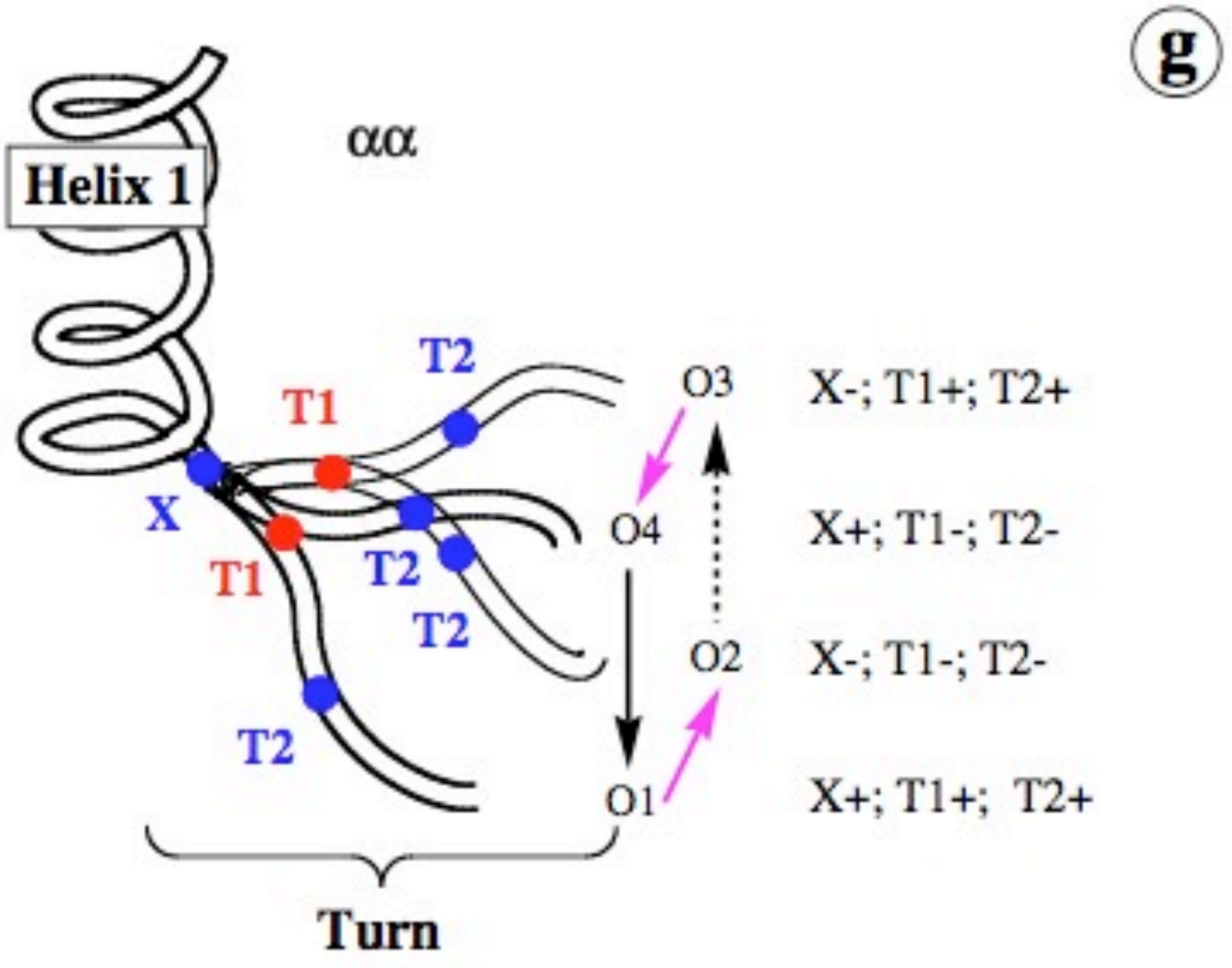

Figure 2

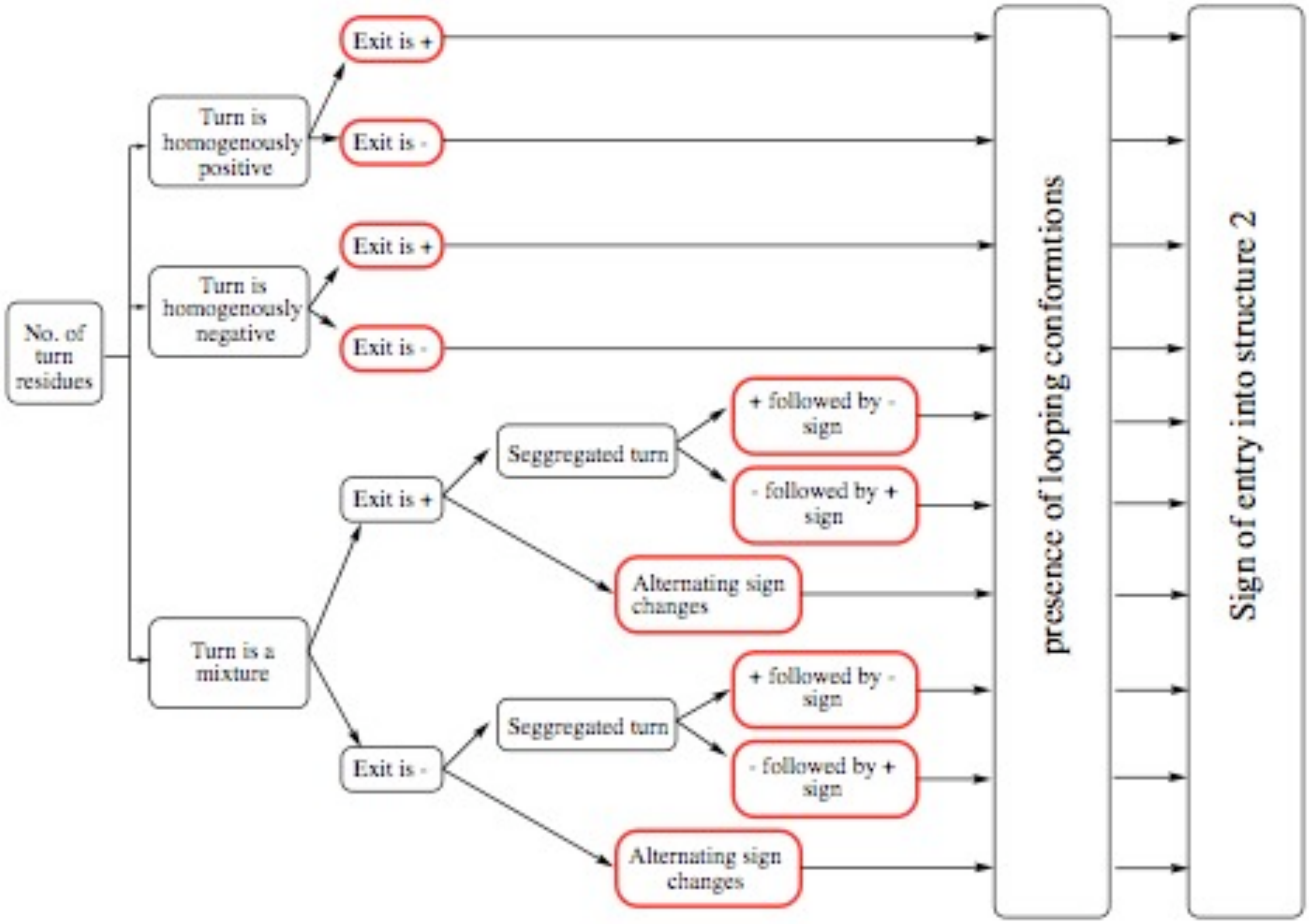
No. of turn residues
Turn is homogenously positive
Turn is homogenously negative
Turn is a mixture
Exit is +
Exit is -
Exit is +
Exit is -
Exit is +
Exit is -
Seggregated turn
Seggregated turn
+ followed by - sign
- followed by + sign
Alternating sign changes
+ followed by - sign
- followed by + sign
Alternating sign changes
presence of looping conformtions
Sign of entry into structure 2

Figure 3

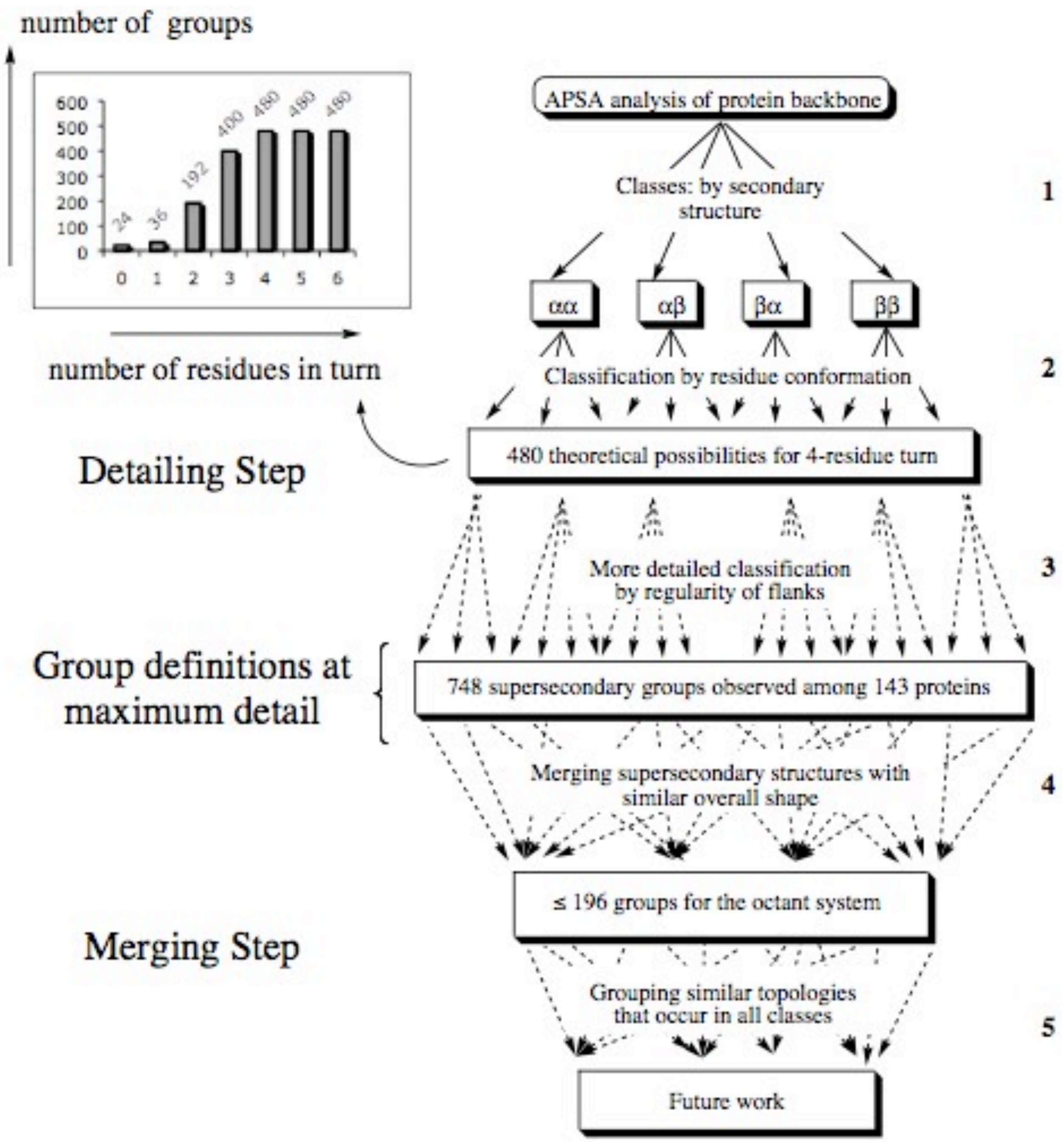

number of groups
600
500
400
300
200
100
0
24
36
192
400
480
480
480
0 1 2 3 4 5 6
number of residues in turn
APSA analysis of protein backbone
Classes: by secondary structure
1
αα
αβ
βα
ββ
Classification by residue conformation
2
Detailing Step
480 theoretical possibilities for 4-residue turn
More detailed classification by regularity of flanks
3
Group definitions at maximum detail
748 supersecondary groups observed among 143 proteins
Merging supersecondary structures with similar overall shape
4
≤ 196 groups for the octant system
Merging Step
Grouping similar topologies that occur in all classes
5
Future work

Figure 4

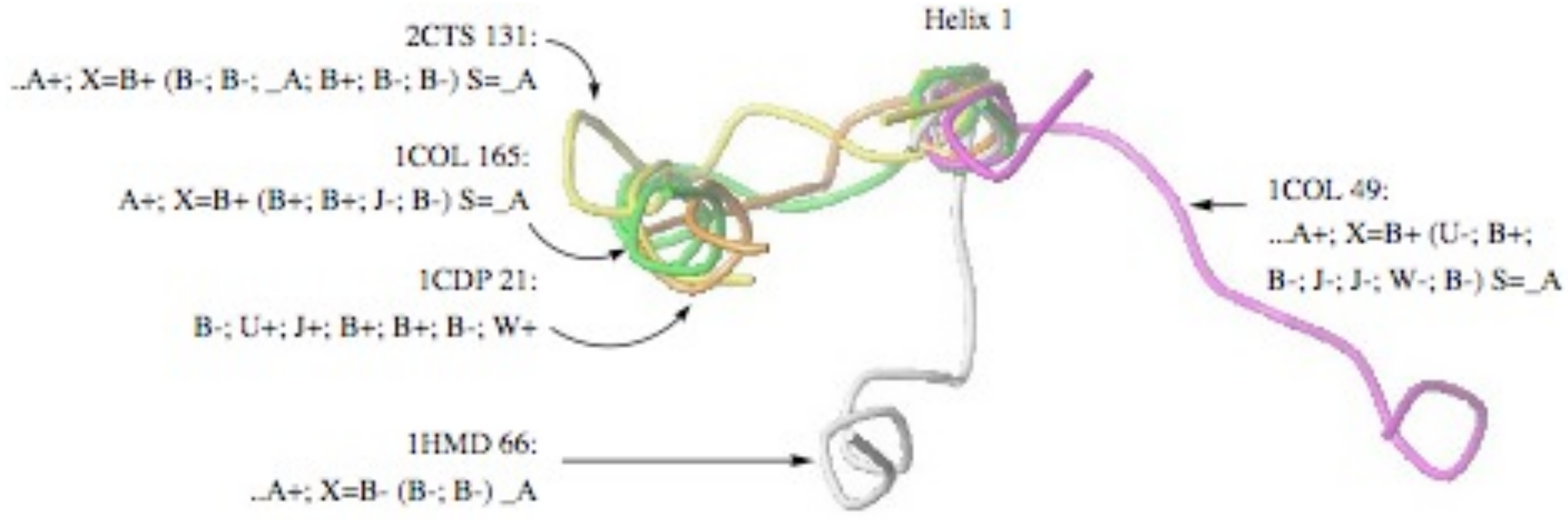
2CTS 131:
..A+; X=B+ (B-; B-; _A; B+; B-; B-) S=_A
1COL 165:
A+; X=B+ (B+; B+; J-; B-) S=_A
1CDP 21:
B-; U+; J+; B+; B+; B-; W+
1HMD 66:
..A+; X=B- (B-; B-) _A
Helix 1
1COL 49:
...A+; X=B+ (U-; B+;
B-; J-; J-; W-; B-) S=_A

Figure 5a, b

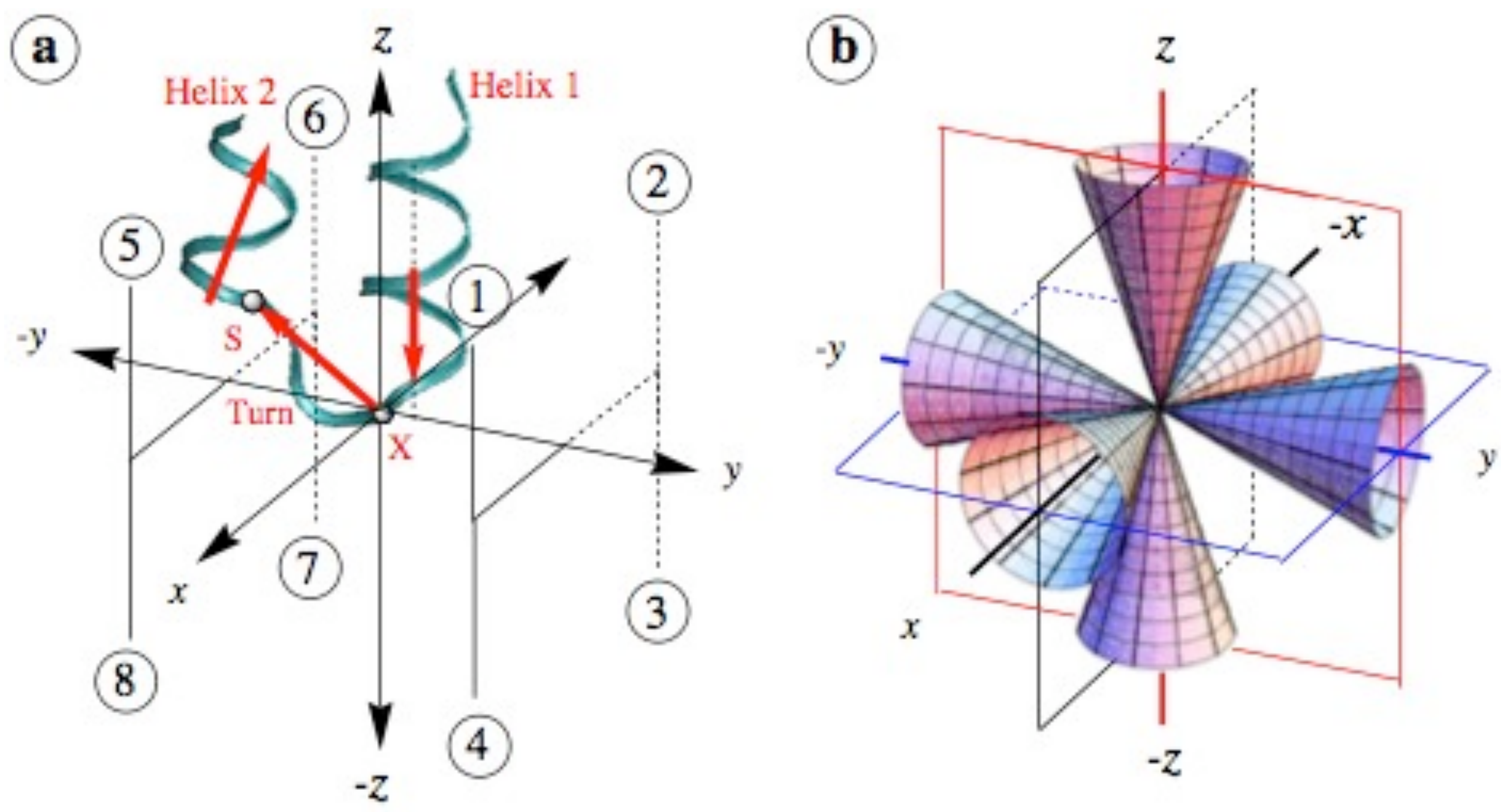

Figure 5c, d

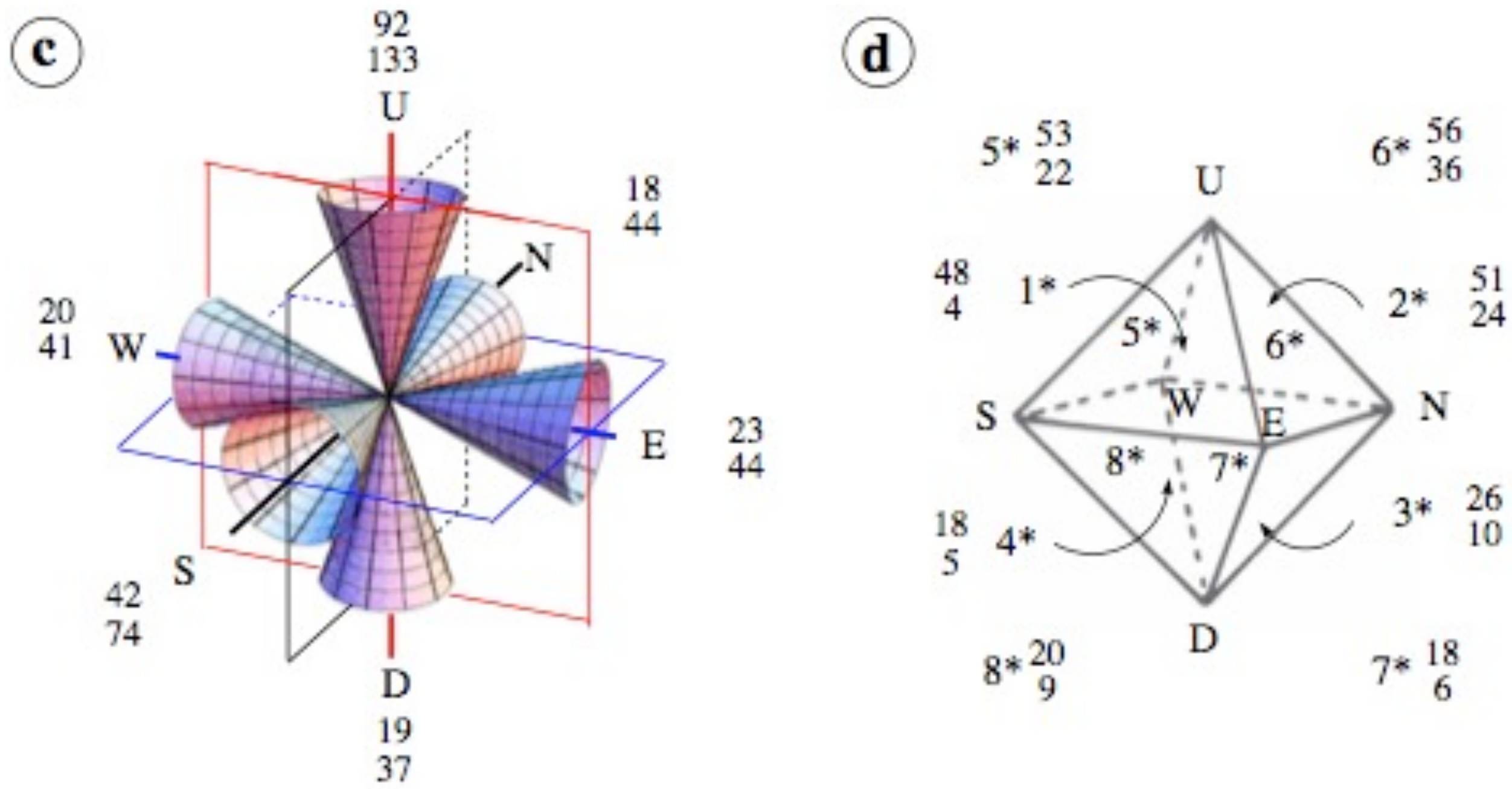